\documentclass[12pt,a4paper]{article}
\usepackage{t1enc}
\usepackage[dvips,final]{graphicx} %<--ked chces vkladat obrazky
\usepackage{tabularx} %<--ked chces robit nieco s tabulkami
\usepackage{amsfonts} 
\usepackage{amsthm} 

\def\du{\unskip\smash{\lower 1.4ex \hbox{\char34}}\kern-.2ex}
\def\hu{\kern-.2ex\hbox{\char92}}

\newcommand{\bdis}{\begin{displaymath}}
\newcommand{\edis}{\end{displaymath}}
\newcommand{\be}{\begin{equation}}
\newcommand{\ee}{\end{equation}}
 
\newcommand{\pd}{\partial} 
\newcommand{\upm}{u^{(\pm)}}  
\newcommand{\um}{u^{(-)}} 
\newcommand{\up}{u^{(+)}} 
\newcommand{\om}{\omega} 
\newcommand{\rmd}{\rm d} 

\newcommand{\mbb}{\mathbb} 
\newcommand{\mcal}{\mathcal} 

\begin{document}
\baselineskip=7.6mm
\newpage

\pagenumbering{roman} 

\title{Casimir Effect in Four Simple Situations - Including a 
Noncommutative Two-Sphere} 
\author{Michal Demetrian\footnote{{\it 
demetrian@sophia.dtp.fmph.uniba.sk}}  \qquad \\ 
Department of Theoretical Physics \\ 
Faculty of Mathematics, Physics and Informatics \\ 
Mlynska Dolina F2, 842 48 Bratislava \\ 
Slovak Republic} 
\maketitle 

\abstract{A computation of the Casimir effect for a real scalar field 
in four situations: on a 
segment of a line, on a circle and on both standard commutative and 
noncommutative two-spheres is given in this paper. The main aim of this 
paper is to discuss the Casimir energy on noncommutative sphere within the 
theory with commutative time. The comparasion with the noncommutative 
cylinder is also done.} 

\section{Introduction} 
The energy of the vacuum - zero-point energy - is a 
direct consequence of the quantum mechanics. Since the birth of quantum 
mechanics a central question has been whether this 
vacuum state energy 
could have physical (measurable) consequences. 
Casimir effect is one of the well-known dynamical effects of the quantum 
vacuum state. Its importance follows from the fact that this effect was 
not only theoretically predicted  
but it was experimentally verified. First time the attraction of two 
parallel neutral conducting plates due to electromagnetic vacuum 
fluctuations  
was measured and then the large set of 
different experiments about the Casimir's idea \cite{cas} 
was done by many 
cooperating experimentators and theoreticians, for more information see 
for example \cite{mak}, \cite{mostep}.  
The vacuum state energy should be of interest from 
the point of view of modern cosmology 
due to production of nonzero cosmological constant \cite{starob}, which can 
drive the inflation process. \\ 
In this paper we are interested in the vacuum state energy of a real 
scalar field $\Phi$ in four situations which differ in a  way in which 
the nonzero vacuum energy is caused. In the second section the scalar 
field on an interval (finite segment of a line) is discussed. The nonzero 
vacuum-state energy appears as the consequence of the boundaries.  
This example is most similar to the original Casimir's idea \cite{cas} 
confirmed by the measurments of attraction of uncharged conducting plates. 
In the third and fourth sections the Casimir's energy of the scalar field 
on a circle (one dimensional sphere) and a two dimensional sphere is 
discussed. In these cases the nontrivial properties of vacuum state follow 
from the nontrivial (different from the Minkowski space) topology. In the 
fifth section the new ideas of noncommutative geometry are implemented: 
Casimir's energy is computed on a two dimensional fuzzy-sphere. In this 
case "the problems" with renormalization and regularization disappear. 
Noncommutativity yields a natural cutoff by introducing a new fundamental 
constant, fundamental in the sence of being similar  
to the Planck constant. 

\section{Scalar field on an interval} 
We start with a real scalar field $\Phi=\Phi(t,x)$ defined on a space 
interval 
of length $a$: $x\in \langle 0,a\rangle$. It means that our spacetime is a 
strip of two dimensional Minkowski space\footnote{We are using the  
notation that space-time interval $ds$ is given by 
$ds^2=c^2dt^2-dx^2=dx_\mu dx^\mu$ 
and $x^0=ct$, $x^1=x$ and greek indices $\in \{ 0,1\}$.} 
with a space size equal to $a$. The dynamics of the classical field $\Phi$ 
with the mass $m$ is given by the Lagrangian density 
\be \label{L} 
\mcal{L}=\frac{1}{2}\pd_\mu\Phi\pd^\mu\Phi 
-\frac{1}{2}\frac{m^2c^2}{\hbar^2}\Phi^2 \quad ,  
\ee 
from which the well-known equation of motion (Klein-Gordon) follows 
\be \label{2} 
\frac{1}{c^2}\frac{\pd^2\Phi}{\pd t^2}-\frac{\pd^2\Phi}{\pd 
x^2}+\frac{m^2c^2}{\hbar^2}\Phi=0 \quad . 
\ee 
Now we are to impose some boundary conditions on the (classical) field 
$\Phi$. We have to be aware of importance of boundary conditions 
which play an important role because they define the situation and should 
be imposed in a physical way in a realistic situation\footnote{For example 
 in the case 
of an electromagnetic field we have the well-known boundary conditions on 
the surface of an ideal conductor.}. We have chosen the Dirichlet 
conditions  
\be \label{3} 
\Phi(t,x=0)=\Phi(t,x=a)=0 \quad . 
\ee 
The complete orthonormal set of functions obeying the boundary problem 
(\ref{2}),(\ref{3}) with respect to the scalar product related to the eq. 
(\ref{2}) 
\be \label{scp}  
(\Phi_1 ,\Phi_2):=\frac{i}{c}\int_0^a {\rmd} x
\left( \Phi_1^*\pd_t\Phi_2-\Phi_2\pd_t\Phi_1^*\right) \quad , 
\ee    
is as follows 
\be \label{4} 
\upm_n(t,x)=\left(\frac{c}{a\om_n}\right)^{1/2} e^{\pm 
i\om_nt}\sin(k_nx) \quad , 
\ee 
where 
\bdis 
\om_n=\left[ \frac{m^2c^4}{\hbar^2}+c^2k_n^2\right]^{1/2} \quad , \qquad 
k_n=\frac{\pi n}{a} \quad , \qquad n\in\{ 1,2,3,\dots\} \quad . 
\edis 

\newcommand{\va}{|0\rangle} 
\newcommand{\ep}{\epsilon}

Any solution to our boundary problem can be expanded into the 
series of  
functions (\ref{4}). The canonical quantization of the field $\Phi$ is  
performed by means of such an expansion: 
\be \label{exp1} 
\Phi(t,x)=\sum_{n=1}^{\infty}
\left[\up_n(t,x)a^+_n+\um_n(t,x)a_n\right] \quad , 
\ee 
where $a, \quad a^+$ are the annihilation and creation operators obeying 
the canonical commutation relations 
\be \label{ccr} 
\left[ a_n,a^+_m\right]=\delta_{nm} \quad , \qquad \left[a_n,a_m\right] = 
\left[a_n^+,a_m^+\right]=0 \quad . 
\ee 
The vacuum state $\va$ 
is specified as usual by the conditions  
\be \label{vac} 
a_n\va =0 \qquad \forall n \quad . 
\ee 
We are interested in the energy of this state, it means we would like to 
compute the vacuum expectation value of the Hamiltonian density $H$ 
\be \label{ham1} 
H=\frac{\pd \mcal{L}}{\pd(\pd_t\Phi)}\pd_t \Phi-\mcal{L}
=\frac{\hbar c}{2}\left[ 
\frac{1}{c^2}(\pd_t\Phi)^2+(\pd_x\Phi)^2\right]+\frac{1}{2}\frac{m^2c^2}
{\hbar^2}\Phi^2 \quad . 
\ee 
Inserting eq. (\ref{exp1}) into (\ref{ham1}) with respect to eqs. 
(\ref{ccr}) and eq. (\ref{4}) we easily get the energy density in the 
form 
\bdis 
\langle 0|H|0\rangle 
=\frac{\hbar}{2a}\sum_{n=1}^{\infty}\om_n-\frac{m^2c^4}{2a\hbar}\sum_{n=1}^
{\infty} \frac{\cos(2k_nx)}{\om_n} \quad , 
\edis  
so, the total energy of the vacuum state $E(a,m)$ 
is the integral over 
$x\in\langle 0,a\rangle$ of the energy density given by the previous 
formula 
\be \label{en1} 
E(a,m)=\frac{\hbar}{2}\sum_{n=1}^{\infty}\om_n \quad . 
\ee 
The quantity $E(a,m)$ is evidently divergent but there is a standard 
possibility to give a meaning to it using a regularization. One of the 
simplest methods used in the authentic work \cite{cas} is to introduce a 
dumping function (something like Boltzman factor) 
$\exp(-\ep\om)$ with $\ep>0$ 
behind the summation 
sign in the eq. (\ref{en1}). We shall start disscusion about the 
regularization with the case of massless field ($m=0$) - it simplifies the 
situation. The regularized energy is 
\be 
E_\ep(a,0)=\frac{\hbar}{2}\sum_{n=1}^{\infty}\frac{c\pi n}{a}\exp
\left(-\frac{c\pi 
n}{a}\ep\right)=\frac{c\pi}{a}\frac{\hbar}{8}\frac{1}
{\sinh^2\left(\frac{c\pi\ep}{2a}\right)} \quad . 
\ee 
Now we can expand the quantity $E_\ep(a,0)$ into the power series in the 
regularization parameter $\ep$ and separate $E_\ep(a,0)$ into a sum of 
two parts: the first - singular in regularization removing ($\ep\to 0$) 
and the second - finit contribution to the energy 
in the same regularization removing. 
The expansion we are talking about looks like 
\be \label{exp2}
E_\ep(a,0)=\frac{c\pi}{a}\frac{\hbar}{8}
\left[ \frac{a^2}{c^2\hbar^2}\frac{4}{\ep^2}-\frac{1}{3}+O(\ep^2)\right] 
\quad . 
\ee 
Let us denote by $E^{phys}(a,0)$ the physical (relevant) energy of our 
vacuum state. Now we would like to identify this quantity with the 
nonsingular term in the eq. (\ref{exp2}), i.e. to write 
\be \label{enphys} 
E^{phys}(a,0)=-\frac{\pi\hbar c}{24a} \quad , 
\ee  
but how to argue that this is the case? One kind of argument could be 
grounded on the following. We are not interested in the energy but in the 
energy differences\footnote{It is always so if the gravitation does not 
play any role.}, 
i.e. $\mbox{Force}=(E(a+{\rmd} a,0)-E(a,0))/{\rmd} a$ is 
significant rather than energy. It means that if we perform several 
regularizations and renormalizations we have to get the same result for 
the force. 
We can choose the energy of Minkowski 
vacuum per length $a$: 
($E^M(a,m)$) to be zero and {\bf define} the physical value of vacuum 
energy by 
\be \label{endef} 
E^{phys}(a,m):=\lim_{\ep\to 0^+}\left[ E_\ep(a,m)-E^M_\ep(a,m)\right] 
\quad . 
\ee 
Let us compute $E^M(a,m)$ 
in the canonical way presented above. The result is 
\be \label{enmink} 
E^M(a,m)
=\frac{\hbar a}{2\pi}\int_0^{\infty}\om{\rmd} k \quad , \quad \mbox{where} 
\quad \om=\left(\frac{m^2c^4}{\hbar^2}+c^2k^2\right)^{1/2} \quad . 
\ee 
It is easy to compute the regularized value $E_\ep^M(a,m)$ for the massless 
field. One gets 
\be 
E^M_\ep(a,0)=\frac{\hbar a}{2\pi}\int_0^{\infty}\om 
e^{-\ep\om}{\rmd} k 
=\frac{\hbar a}{2\pi c}\int_0^\infty\om e^{-\ep\om}{\rmd \om} 
=\frac{\hbar a}{2\pi c}\frac{1}{\ep^2} \quad . 
\ee 
Using the definition formula (\ref{endef}) for the physical value of 
vacuum state energy we instantly get above expectated result 
(\ref{enphys}). \\   
Macroscopic effect of the vacuum energy is the force (attractive) 
between the end 
points of the interval 
given in account to the eq. (\ref{enphys}) by 
\bdis 
F=-\frac{{\rmd} E^{phys}(a,0)}{{\rmd} a}=-\frac{\pi\hbar c}{24a^2} \quad . 
\edis

\newcommand{\la}{\lambda} 
\newcommand{\La}{\Lambda}

Now we shall rewrite the above described alghoritm  
into more convenient and usuable form. The 
renormalized vacuum energy (\ref{endef}) is of the form 
\bdis 
\lim_{\ep\to 0^+}
\left[ \sum_{n=0}^{\infty}F_\ep(n)-\int_0^\infty F_\ep(x){\rmd} x \right] 
\quad , 
\edis 
where $F_\ep(n)=\frac{\hbar}{2}\frac{c\pi n}{a}f(\ep\om)$, with 
$f(\ep\om)$ - (smooth) dumping function, and $F$ is 
an analytic function in the complex $z$ half-plane $Re(z)>0$ and the sum 
and integral exist. Than we can use the so-called Abel-Plana formula 
(see eg. \cite{je}) 
\be \label{ap1} 
\sum_{n=0}^{\infty}F_\ep(n)-\int_0^\infty F_\ep(x){\rmd} x = 
\frac{1}{2}F_\ep(0)+i\int_0^\infty\frac{F_\ep(it)-F_\ep(-it)}{e^{2\pi t}-1} 
{\rmd} t \quad . 
\ee 
The integral on the right-hand side of the eq. (\ref{ap1}) converges 
uniformly for all $\ep>0$, so we are allowed to perform the limit 
before the integration. We easily get 
\begin{eqnarray}  
E^{phys}(a,0)& = & \frac{1}{2}0+i\int_0^\infty 
\frac{\frac{\hbar}{2}\frac{c\pi}{a}(it)-\frac{\hbar}{2}\frac{c\pi}{a}(-it)}
{e^{2\pi t}-1}{\rmd} t = -\frac{\hbar c\pi}{a}\int_0^\infty\frac{t{\rmd } 
t}{e^{2\pi t}-1} \nonumber \\ 
& = & -\frac{\hbar c\pi}{a}\frac{1}{(2\pi)^2}
\sum_{k=1}^{\infty}\int_0^\infty ue^{-ku}{\rmd} u =-\frac{\hbar 
c\pi}{a}\frac{1}{(2\pi)^2}\frac{\pi^2}{6} \nonumber \\ 
& = & -\frac{\hbar c \pi}{24 a} 
\quad 
, 
\end{eqnarray} 
that is in acoord with (\ref{enphys}). \\ 
Another way how to obtain the result (\ref{enphys}) is to use the 
renormalization that uses the so-called zeta function regularization (see 
eg. \cite{bs}). The idea is to take the formula (\ref{en1}) and to see 
that it can be formally understood as follows 
\be \label{zeen} 
E(a,0)=\frac{\hbar}{2}\frac{c\pi}{a}\sum_{n=1}^\infty n 
=:\frac{\hbar}{2}\frac{c\pi}{a}\zeta(-1)=E^{phys}(a,0) \quad , 
\ee  
where $\zeta$ is the Riemann zeta function defined for complex $z$ with 
$Re(z)>1$ by the sum  
\bdis 
\zeta(z)=\sum_{n=1}^{\infty}n^{-z} \quad . 
\edis 
The expression $\zeta(-1)$ has to be understood as the value of analytic 
continuation (see eg. \cite{je}) of the Riemann zeta function into a whole 
complex plane. Such a continuation is given by the functional 
formula\footnote{$\Gamma(z)$  is the Euler's gamma function.}  
(\cite{je} or \cite{rg}) 
\bdis 
\Gamma\left( \frac{z}{2}\right) \pi^{-z/2}\zeta(z)=\Gamma\left( 
\frac{1-z}{2}\right)\pi^{(1-z)/2}\zeta(1-z) \quad . 
\edis 
Using previous formula we get $\zeta(-1)=-1/12$, so substituiting this 
value into the eq. (\ref{zeen}) we get instantly the result 
(\ref{enphys}). \\ 
Now we are going to use the procedure of renormalization grounded on the 
Abel-Plana formula (\ref{ap1}) for the massive scalar field $\Phi$. We 
have (\ref{en1}) and (\ref{4}) 
\bdis 
E_\ep(a,m)=\frac{\hbar}{2}\sum_{n=1}^\infty \om_n 
e^{-\ep\om_n}=-\frac{mc^2}{4}+\sum_{n=0}^{\infty}\om_ne^{-\ep\om_n} \quad 
. 
\edis 
The above calculations with 
$F_\ep(x)=\hbar/2\sqrt{m^2c^4/\hbar^2+\pi^2c^2/a^2x^2}$ lead to the result 
\begin{eqnarray}  
E^{phys}(a,m) & = &  
-\frac{mc^2}{4}-\frac{\hbar c}{4\pi a}\int_\la^\infty 
\frac{\sqrt{t^2-\la^2}}{e^{t}-1}{\rmd}t=-\frac{mc^2}{4} \nonumber \\ 
& - & 
\frac{\hbar c}{4\pi a}\la^2\int_1^\infty 
\frac{\sqrt{t^2-1}}{e^{\la t}-1}{\rmd}t \nonumber \\   
& = &  -\frac{mc^2}{4}-\frac{\hbar c}{4\pi 
a}\la^2\int_0^\infty\frac{\sinh^2(u)}{e^{\la \cosh(u)}-1}{\rmd} u 
\quad , 
\end{eqnarray} 
where 
\bdis 
\la=2\frac{mca}{\hbar} \quad . 
\edis 
We mention that the additive constant term $-mc^2/4$ does not contribute 
to the force and that putting $m=0$ we obtain the eq. (12). One can 
investigate the behaviour of $E^{phys}(a,m)$ in the limit of very massive  
field ($\la\to\infty$). It is easy to see that the Casimir's energy and 
corresponding Casimir's force are 
exponentially ($\sim e^{-\la}$) suppressed in this case. Let us 
notice that $\la$ is nothing else as double length of our interval 
measured in the unit  
of the Compton wave-length of matching particle $\Phi$.

\section{Scalar field on a circle} 

Let us consider now the scalar field $\Phi$ on a circle with radius 
equal to $a>0$. It means that our spacetime is a cylinder of radius $a$ 
($x$ - space coordinate: $x\in (0,2\pi a)$ ) 
with time belonging to $(-\infty,\infty)$. Equation of 
motion for such a field is Klein-Gordon equation (\ref{2}). The case now 
differs from the previous one by the boundary conditions imposed on the 
field. Topology of the circle defines the periodic boundary conditions 
\be \begin{array}{ccc} 
\Phi(t,0) & = & \Phi(t,2\pi a) \quad , \\ 
\pd_x\Phi(t,0) & = & \pd_x\Phi(t,2\pi a) \quad .  
\end{array} 
\ee 
Canonical quantization leads to the expansion of the field $\Phi$ given by 
the eq. (\ref{exp1}) with the operators $a_n$ and $a^+_n$ obeying 
canonical commutation relations (\ref{ccr}). 
Normalized mode functions $\upm_n$ are 
given by the formulae  
\begin{eqnarray}  \label{exp3} 
& & 
\upm_n(t,x)=\left(\frac{c}{2a\om_n}\right)^{1/2}
\exp(\pm i(\om_nt-k_nx)) \quad , 
\nonumber \\ & &  
\om_n=\left[ \frac{m^2c^4}{\hbar^2}+c^2k_n^2\right]^{1/2} 
\quad \mbox{and} \quad 
k_n=\frac{n}{a} \quad \mbox{with} \quad n\in \mbb{Z} \quad . 
\end{eqnarray} 
Using the expansion (\ref{exp1}) with respect to the formulae (\ref{exp3}) 
we can get the unrenormalized energy of the vacuum (once again 
specified by (\ref{vac})) in the form 
\be 
E(a,m)=\int_0^{2\pi a}\langle 0|H(t,x)|0\rangle{\rmd} 
x=\frac{\hbar}{2} 
\sum_{n=-\infty}^{+\infty}\om_n=-\frac{mc^2}{2}+\hbar\sum_{n=0}^{\infty}
\om_n \quad . 
\ee 
The sum expressing $E(a,m)$ is again divergent and should be renormalized. 
We are going to renormalize\footnote{The meaning of the definition formula 
(\ref{endef}) in this case is the same as in the section 2 because the 
Minkowskian vacuum now corresponds to the same one - vacuum of $\Phi$ in  
the two dimensional Minkowski spacetime.} 
it using the Abel-Plana formula (\ref{ap1}) with 
\be  
F(z)=\frac{\hbar c}{a}\sqrt{\la^2 + z^2} \quad \mbox{where} \quad 
\la=\frac{mca}{\hbar} \quad . 
\ee 
The result is 
\be \label{fi2}   
E^{phys}(a,m) = -\frac{2\hbar 
c}{a}\int_\la^\infty\frac{\sqrt{t^2-\la^2}}{e^{2\pi t}-1}{\rmd} t=
-\frac{2\hbar 
c}{a}\la^2\int_0^\infty\frac{\sinh^2(u)}{e^{2\pi\la\cosh(u)}-1}{\rmd} u 
\quad . 
\ee 
The special case is the massless field, which correspond to the situation 
$\la$ being zero. In this case we are able to express $E^{phys}$ in an 
algebraical form 
\be \label{fi21} 
E^{phys}(a,0)=-\frac{\hbar c}{12a} \quad . 
\ee 
We note that at $\la\gg 1$  
both Casimir's energy and force are again 
exponentially suppressed by the factor $e^{-\la}$. This follows from the 
eq. (\ref{fi2}).

\begin{figure}[t] 
\includegraphics[width=7cm,height=6cm]{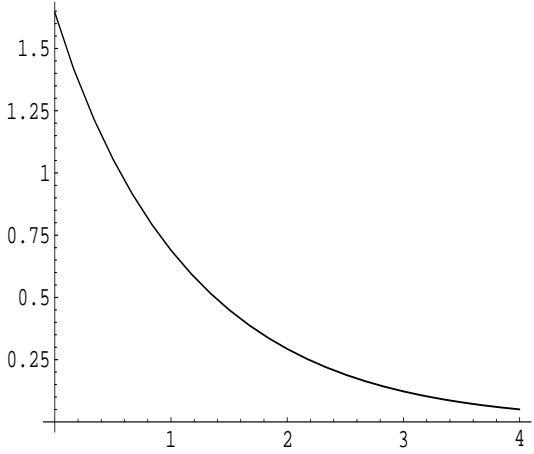} 
\caption{A typical dependence of the Casimir's energy of a scalar field on 
both line segment and circle on $\la$ parameter - plot of the function 
$f(\la)=\int_\la^\infty\sqrt{x^2-\la^2}/(\exp(x)-1)$.}   
\end{figure}

\section{Scalar field on a two dimensional sphere} 
In this section we shall consider a real scalar field on a two dimensional 
sphere ($S^2$) of the radius $a$.  
It means that our spacetime is $S^2\times \mbb{R}^1$. Now 
it is a little bit 
more difficult to give the dynamics of the field as in the 
previous two sections. It is so due to 
the fact that our current spacetime is 
curved (its Riemann curvature tensor is not zero). In general 
a lot of uglinesses  
follow from this fact  at quantum level (see eg. \cite{bd}). 
Our point of view will be 
canonical. We define the dynamics of the field $\Phi$ giving the 
Lagrangian density 
\be \label{las} 
\mcal{L}=\frac{1}{2}\sqrt{|g|}
\left[ g^{\mu\nu}\pd_\mu\Phi\pd_\nu\Phi-[m^2-\Xi  
R]\Phi^2\right] \quad , 
\ee 
where $g_{\mu\nu}$ with are the components of the metric, in the 
standard spherical coordinates $(\theta,\phi)$ one has 
\bdis 
g=c^2dt\otimes dt-a^2
\left[d\theta\otimes d\theta+\sin^2(\theta)d\phi\otimes 
d\phi\right] \equiv g_{\mu\nu}dx^\mu\otimes dx^\nu \quad , 
\edis 
where $|g|$ is the absolute value of the determinant of 
metric, $m$ is the mass 
of the field, $R$ is the scalar curvature (computed with respect to the 
Levi-Civita conection - \cite{bd}) and $\Xi$ is a coupling constant to the 
scalar curvature - at the moment $\Xi$ is not specified. \\ 
We shall use the coordinates $(t,\theta,\phi)$ in what follows, so we put 
down: $g=ca^2\sin(\theta)$. The scalar curvarure is given by 
\bdis R=-\frac{2}{a^2} \quad . \edis 
Equation of motion which follows 
from the Lagrangian density (\ref{las}) is the covariant wave equation 
\be 
|g|^{-1/2}\pd_\mu\left(|g|^{1/2}g^{\mu\nu}\pd_\nu\Phi\right)+(m^2-\Xi 
R)\Phi=0 \quad , 
\ee 
or in the coordinates one obtains 
\be \label{kgs} 
\frac{\pd^2\Phi}{\pd t^2}-\frac{c^2}{a^2}\Delta_{\theta\phi}\Phi+c^2
\left( \frac{m^2c^2}{\hbar^2}+\frac{2\Xi}{a^2}\right)\Phi=0 \quad , 
\ee 
where $\Delta_{\theta\phi}$ is the standard Laplace operator on $S^2$. \\ 
We see that nonzero value of $\Xi$ corresponds only to the redefinition of 
the mass. It is so due to 
the independence of the metric on the time. In the 
general case the coupling to the curvature is very important and leads to 
many effects  - see \cite{bd}. We shall use the so-called conformal 
coupling $\Xi=1/8$. We can 
introduce the new (effective) mass $M$ as follows 
\be \label{massM} 
\frac{M^2c^2}{\hbar^2}:=\frac{m^2c^2}{\hbar^2}+\frac{2\Xi}{a^2} \quad . 
\ee 
Now we have to find a complet orthonormal set of solutions to the eq. 
(\ref{kgs}) - say $\upm(t,\theta,\phi)$. One can separate variables in the 
eq. (\ref{kgs}) putting 
\bdis 
\Phi(t,\theta,\phi)=\exp(i\om t)Y_{lm}(\theta,\phi) \quad , 
\edis 
where $Y_{lm}$ are the spherical harmonics - they are eigenfunctions 
of the Laplace operator $\Delta_{\theta\phi}$ on the sphere with  
eigenvalues $-l(l+1)$ where $l\in\{ 0,1,2,\dots\}$ and $m\in\{ 
-l,-l+1,\dots ,l\}$. Substituiting this ansatz to the eq. (\ref{kgs}) we 
get the following set $\upm$ of solutions to our equation 
\be \label{upms} \begin{array}{ccc}  
\up_{lm}(t,\theta,\phi) & = & 
\frac{1}{a}\sqrt{\frac{c}{2\om_l}}\exp(i\om_lt)Y_{
lm}(\theta,\phi) \quad , \\ 
\um_{lm}(t,\theta,\phi) & = & (\up_{lm}(t,\theta,\phi))^* \quad , 
\end{array} \ee 
which is orthonormal (with respect to the scalar product (\ref{scp})) and 
complete (this follows from the properties of spherical harmonics). The 
frequency $\om_l$ (as well as energy) depends only on the quantum number 
$l$ and is given by 
\be \label{freqs}  
\om_l=
\left[\frac{m^2c^4}{\hbar^2}+\frac{c^2}{a^2}(2\Xi+l(l+1))\right]^{1/2}=
\left[\frac{M^2c^4}{\hbar^2}+\frac{c^2}{a^2}l(l+1)\right]^{1/2} \quad . 
\ee  
Field operator $\Phi$ can be expanded into the modes $\upm_{lm}$ as 
follows 
\be \label{exp11} 
\Phi=\sum_{l=0}^{\infty}\sum_{m=-l}^{l}
\left[\up_{lm}a^+_{lm}+\um_{lm}a_{lm}\right] \quad , 
\ee  
where $a^+$ and $a$ are creation and annihilation operators obeying the 
commutation relations 
\bdis 
[a^+_{lm},a_{l'm'}]=-\delta_{ll'}\delta_{mm'} \quad 
[a_{lm},a_{l'm'}]=[a^+_{lm},a^+_{l'm'}]=0 \quad . 
\edis 
Now we can insert this expansion into the Hamiltonian density,   
constructed from the Lagrangian density (\ref{las}) in a canonical way, 
integrate over all sphere with the result - unrenormalized energy of the 
vacuum state is\footnote{In this formula this property of spherical 
harmonics has been used: $$ \sum_{m=-l}^{l}Y_{lm}(\theta,\phi)
Y^*_{lm}(\theta,\phi)=\frac{2l+1}{4\pi}$$} 
\begin{eqnarray} \label{esp} 
E(a,m) & = & \frac{\hbar}{2}\sum_{l=0}^{\infty}(2l+1)\om_l =
\hbar\sum_{l=0}^\infty\left( l+\frac{1}{2}\right)\om_l = \nonumber \\ 
& = &  
\hbar\sum_{l=0}^\infty\left( l+\frac{1}{2}\right)
\sqrt{\frac{M^2c^4}{\hbar^2}+\frac{c^2}{a^2}l(l+1)} \nonumber \\ 
& = & \frac{\hbar c}{a}\sum_{l=0}^\infty\left( l+\frac{1}{2}\right)
\sqrt{\frac{m^2c^2}{\hbar^2}+\left( l+\frac{1}{2}\right)^2} \quad . 
\end{eqnarray} 
Since $E(a,m)$ is divergent it should be renormalized. We are going to use 
the Abel-Plana formula (its light modification) 
to do it. But before do this we would like to mention 
that the fact $M^2\geq 0$ follows 
from the formulae (\ref{upms}) and (\ref{freqs}) and its physical 
interpretation. If it is not the case we will get at least one 
exponentially (in time) 
expanding and decreasing mode that does not play with the probability 
preservation. One can understant to this fact as a constraint to the 
value of $\Xi$ at the given values of $m$ and $a$. \\ 
The possibility to use something like Abel-Plana formula (\ref{ap1}) 
is grounded on the fact (we are using the same logic as in the second 
section now) that the energy of Minkowski vacuum (2+1 Minkowski space) per 
area of the sphere of radius $a$ ($4\pi a^2$) is given by 
\begin{eqnarray}
E^{M}(a,m)& = & 
4\pi 
a^2\frac{\hbar}{2}\int_{-\infty}^\infty\int_{-\infty}^\infty\frac{{\rmd} 
k_1{\rmd}
k_2}{(2\pi)^2}c\sqrt{k_1^2+k_2^2+\frac{m^2c^2}{\hbar^2}}= \nonumber \\ 
& & a^2\hbar 
c\int_0^\infty{\rmd} k k\sqrt{k^2+\frac{m^2c^2}{\hbar^2}} 
 =  \frac{\hbar c}{a}\int_0^\infty{\rmd} z z
\sqrt{z^2+\frac{M^2c^2a^2}{\hbar^2}} \quad . 
\end{eqnarray} 
Now, combining the last two formulae, 
we are able to use the half-integer Abel-Plana formula (see 
\cite{je}): 
\bdis 
\sum_{n=0}^{\infty}F(n+1/2)-\int_0^\infty{\rmd} z 
F(z)=-i\int_0^\infty\frac{F(iz)-F-iz)}{e^{2\pi z}+1} \quad , 
\edis 
with $F(z)=\hbar c/a z\sqrt{z^2+\la^2}$ where 
\bdis \la=\frac{mca}{\hbar} \quad , \edis 
to get the final result for the Casimir's energy. Computations lead to 
\be \label{sphres} 
E^{phys}(a,m)=2mc^2\la^2\int_0^1\frac{z\sqrt{1-z^2}}{\exp(2\pi\la 
z)+1}{\rmd} z \quad . 
\ee 
\begin{figure}[h] 
\includegraphics[width=7cm,height=6cm]{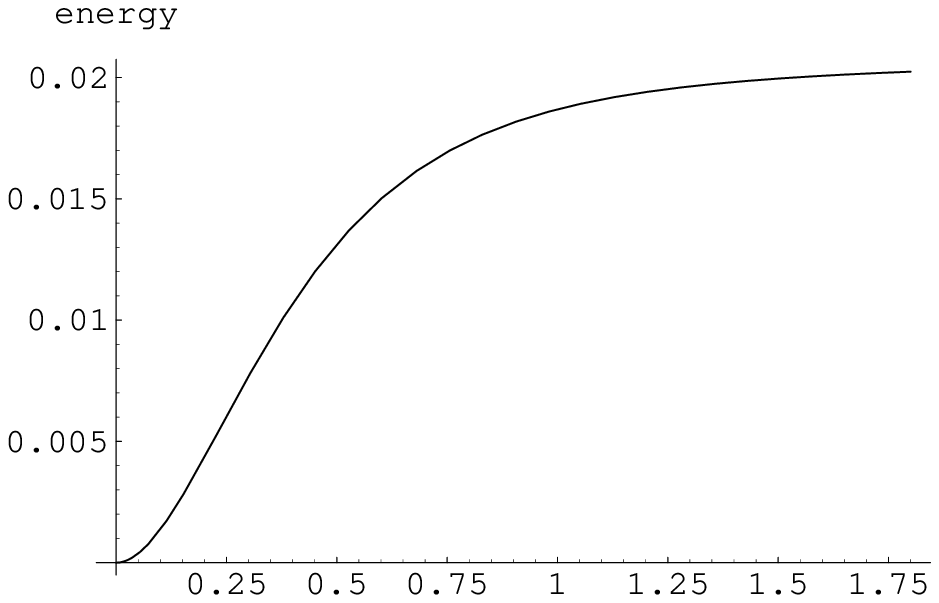} 
\includegraphics[width=7cm,height=6cm]{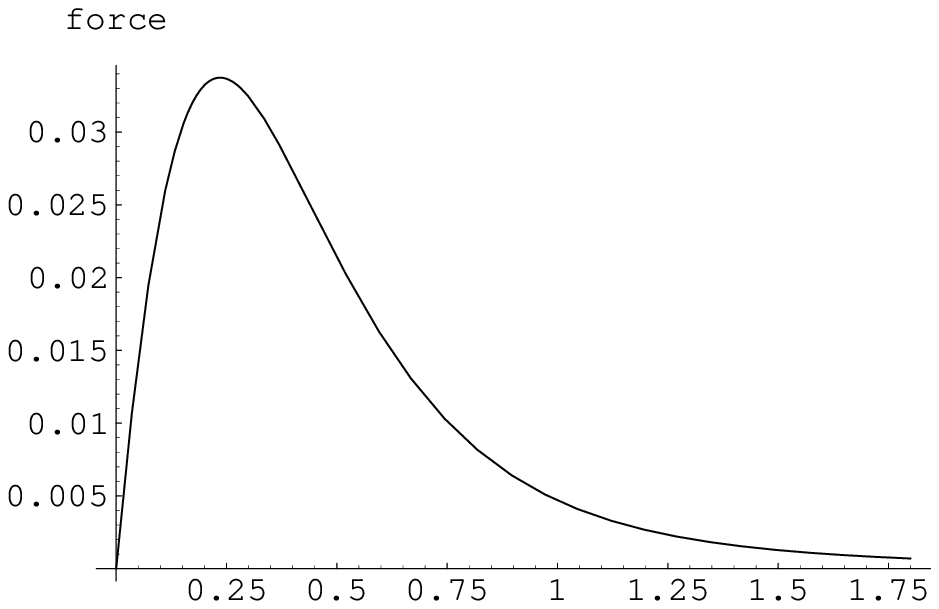}
\caption{Casimir's energy and force 
on the two dimensional sphere. Plots of the functions: 
$f(\la)=\la^2\int_0^1z\sqrt{1-z^2}/(\exp(2\pi \la z)+1){\rmd} z$ (energy) 
and its first derivative (absolute value of the force).} 
\end{figure} 

The Casimir's energy on a sphere is not exponentially suppressed 
at large 
value of the field mass $m$, but the force is. So we have got the similar 
result for the behaviour of the Casimir effect at the very large value of 
the field mass as in the two previous sections. The energy $E^{phys}(a,m)$ 
increases as the radius increases, so the Casimir's force makes the sphere 
be small - it is attractive.

\newcommand{\hx}{\hat{x}} 
\newcommand{\hX}{\hat{X}} 
\newcommand{\ga}{\gamma} 
\newcommand{\hP}{\hat{\Phi}} 
\newcommand{\hJ}{\hat{J}} 
\newcommand{\hY}{\hat{Y}} 

\section{Scalar field on non-commutative two dimensional sphere} 

Since the works of Alan Connes \cite{con} the ideas of the non-commutative 
(NC) geometry have been applied to physics many times. In the case of two 
dimensional models (two space dimensions + external commutative time or 
one space dimension and one noncommuting time) there is a lot of papers in 
which the quantum mechanics and the field theory have been formulated. For 
the review see \cite{other} and the references therein. Special effects 
have been investigated in \cite{speef}. The physical idea to replace our  
standard point of view on the spacetime 
by the NC geometry is based on the statement that at very 
short distances the standard concept of a smooth manifold would change 
into somewhat else. "The short distances" are usually accepted to be of 
the Planck length $l_P$: 
\bdis 
l_P=\left(\frac{\kappa \hbar}{c}\right)^{1/2}\approx 10^{-35}m \quad , 
\edis  
where $\kappa\approx 6.67\times 10^{-11}Nm^2kg^{-2}$ 
is the Newton gravitational constant. One of the main goals 
of the field theories formulated on noncommutative versions of the standard 
spaces is that the UV-regularization appears automatically in such 
theories if the space under consideration is compact (like $S^2$). This is 
caused by the fact that all operators have finite dimensional 
representations (they can be written as the matrices of finite rank) and 
therefore there is no place for UV-divergences. But in the case of 
non-compact spaces the situation quite differs and UV-divergences persist 
in the theory \cite{pers}. \\ 
As we have seen in the section four, the scalar field on the standard - 
commutative - sphere can be expanded into the series (\ref{exp11}), where 
the space-dependence of $\Phi$ is encoded in the spherical harmonics 
$Y_{lm}(\theta,\phi)$. So, at the fixed moment of time it holds 
\bdis 
\Phi(t,\vec{x})=\Phi(\vec{x})=\sum_{lm}\alpha_{lm}Y_{lm}(\vec{x}) \quad , 
\edis 
where $\alpha_{lm}$ are the complex constants obeying:  
$\alpha_{l-m}=\alpha^*_{lm}$ due to the fact the field is real. 
It means that the scalar field $\Phi$ is a 
function $S^2\rightarrow \mbb{R}$ at the fixed moment of time. 
In \cite{tow} the idea is used that the 
property $\Phi$ is defined on $S^2$ is encoded in the properties of 
spherical harmonics $Y_{lm}$ that are a functional representation of the 
rotation group $SO(3)$. The function $\Phi$ can be also treated as the 
function of three real parameters $(x_1,x_2,x_3)\in\mbb{R}^3$ with the 
constraint $x_1^2+x_2^2+x_3^2=a^2>0$. The set of all those functions forms 
the commutative algebra $\mcal{A}_\infty$\footnote{Algebra is a linear 
space where a product ".": 
$\mcal{A}_\infty\times\mcal{A}_\infty\rightarrow\mcal{A}_\infty$ is 
defined and some standard axioms hold. In our case, the algebra's product 
is nothing else as the pointwise product of real-valued functions: 
($\Phi_1.\Phi_2)(x):=\Phi_1(x)\Phi_2(x)$, which is, of course, 
commutative: $\Phi_1.\Phi_2=\Phi_2.\Phi_1$ for each pair of functions 
$\Phi_1,\Phi_2$.}. \\ 
The noncommutativity of the sphere is introduced putting down the 
nontrivial commutation relations 
\be \label{ntcr} 
[\hx_i,\hx_j]=i\ga\ep_{ijk}\hx_k \quad i,j,k\in\{1,2,3\} \quad , 
\ee 
where $\ga$ is a real fundamental constant characterizing the space 
non-commutativity and operators $\hx_i$ supply the standard cartesian 
coordinates $x_i$. 
The constraint defining the sphere radius 
\be \label{rad} 
\hx_1^2+\hx_2^2+\hx_3^2=a^2Id \quad , 
\ee 
is considered, too\footnote{We mention if we considered the relations 
(\ref{ntcr}) we would not straighforward get the NC version of three 
dimensional space (the standard configuration space for a free particle). 
It is so due to the fact that the relations (\ref{ntcr}) are not invariant 
under space translations which are usually considered to be the 
fundamental symmetries of the space.}. 
It is easy to verify that these two relations are not 
in contradiction. Now we loose the interpretation of $\hx_i$ as the points 
due to nontrivial commutation relations (\ref{ntcr}). This situation is 
well-known from the phase-space picture of quantum mechanics where the 
Heisenberg uncertainty principle, as a consequence of commutation relation 
$[x,p_x]=i\hbar$, removes the idea of pointwise phase space of the 
classical mechanics. There are 
no points but only the cells of area $2\pi\hbar$ in which a particle can 
be localized. 
By the scalar field $\hP$ on the NC sphere we shall consider any 
(operator-valued) function of $(\hx_1,\hx_2,\hx_3)$ like in commutative 
case. The set of all those functions forms again an algebra but now the 
algebra is non-commutative due to the eq. (\ref{ntcr}). 
We denote it by $\mcal{A}_N$, 
where a number $N$ should be specified by construction of a representation 
of our commutation relation. \\ 
In \cite{tow} the representation of the commutation relations (\ref{ntcr}) 
with the constraint (\ref{rad}) has been done using Wigner-Jordan 
realization of the generators $\hx_i$ of our NC algebra. The result is 
that such a representation is finite dimensional with dimension of $N$ 
that depends on the values of $a$ and $\ga$ by 
\be \label{N} 
\frac{a}{\ga}=\sqrt{\frac{N}{2}\left(\frac{N}{2}+1\right)} \quad . 
\ee 
Some differential operators like Laplacian 
play an important role in the field theory, especially 
on the sphere. In our 
case we should be interested in Laplacian (squared angular momentum). On 
the standard sphere $\Delta=\sum_{i=1}^{3}J_i^2$, where $J_i$ are the 
anglular momentum operator which may be treated as the Killing's vector 
fields on sphere. They act on a function $\Phi\in \mcal{A}_\infty$ as 
follows 
\bdis 
(J_i\Phi)(\vec{x})=-i\ep_{ijk}x_j\frac{\pd\Phi}{\pd x_k}(\vec{x}) \quad . 
\edis 
There are natural analogues $\hJ_i$
of the operators $J_i$ in the case of NC algebra $\mcal{A}_N$: 
\be \label{nckill} 
\hJ_j\hP:=[\hX_i,\hP] \quad, \mbox{where} \quad \hX_i=\hx_i/\ga \quad 
\mbox{and} \quad \hP\in \mcal{A}_N \quad . 
\ee 
Operators $\hJ_i$ satisfy the same commutation relations as $J_i$ ($su(2)$ 
- commutation relation): $[\hJ_i,\hJ_j]=i\ep_{ijk}\hJ_k$. In \cite{tow} 
the eigenfuctions $\hY_{lm}$ 
of $\sum_i\hJ_i^2=:\hJ^2$ are constructed. They are 
similar to the spherical harmonics, it holds 
\bdis 
\hJ^2\hY_{lm}=l(l+1)\hY_{lm} \quad \mbox{with} \quad l\in\{ 
0,1,2,\dots,N(a,\ga)\} 
\quad m\in\{ -l,\dots ,l\} 
\quad . 
\edis 
The number $N(a,\ga)$ is a natural cutoff. 
Now we could add to our algebra $\mcal{A}_N$ one external 
commutative parameter - time - in accord with the expansion (\ref{exp11}). 
Then the field expansion on the NC sphere with added commutative time is 
of the form 
\be \label{exp111} 
\hP=\sum_{l=0}^{N(a,\ga)}\sum_{m=-l}^{l}
\left[ e^{i\om_lt}\hY_{lm}a^+_{lm}+e^{-i\om_lt}\hY^*_{lm}a_{lm}\right] 
\quad , 
\ee 
where $N(a,\ga)$ is given by (\ref{N}). So, if we omit all no-interesting 
parameters now (like mass, speed of light), we can write for the energy of 
ground state - vacuum $E^{NC}(a,\ga)$ 
\be \label{ncen} 
E^{NC}(a,\ga)\sim 
a^2\sum_{l=0}^{N(a,\ga)}(2l+1)\sqrt{1+\frac{1}{a^2}l(l+1)} 
\quad . 
\ee 
We see that $E^{NC}(a,\ga)$ is finite, so no regularization is needed. 
There is shown a dependence of the Casimir's energy (\ref{ncen}) on the 
spere radius $a$ on the two 
figures bellow. For large value of $N$, i.e. large 
value of the fraction $a/\ga$ it is almost proportional to $a^4$, so the 
Casimir's force, which is also attractive, is proportional to $a^3$. At the 
small values of the fraction $a/\ga$ this dependence is stair-step but 
always Casimir energy increases with the sphere radius.  
So we see that the result is not in accord with the commutative 
one where the Casimir's force might achieve a maximum at the suitable 
combination of parameters $a$ and $m$. 
\begin{figure}[h] 
\includegraphics[width=7cm,height=6cm]{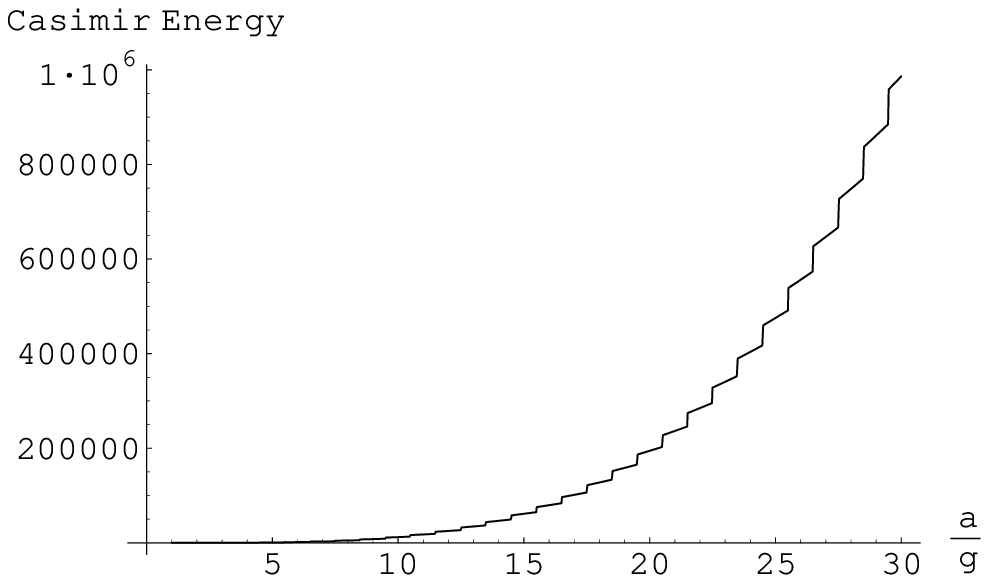} 
\includegraphics[width=7cm,height=6cm]{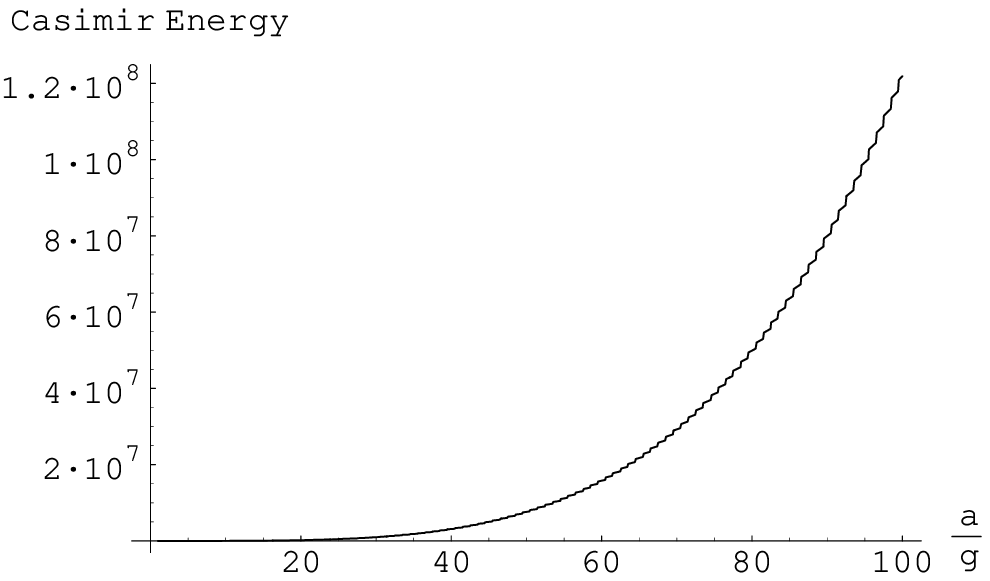} 
\caption{Casimir's energy on a noncommutative two-sphere.} 
\end{figure} 

\section{Discussion} 

In this paper, we have investigated the Casimir effect in various 
situations. In the sections 2-4 the computations of Casimir's energy of a 
scalar field on the standard smooth spaces have been done. Two different 
ways of appearing of Casimir effect have been presented: in the first 
section the Casimir's energy is caused by the boundaries and in the 
sections two and three by the topologies of spaces under consideration. 
These results are known 
and may be found in large literature, for example see the references 
\cite{mak}, \cite{bd} and the references therein. The new result is 
presented in the fifth section where the Casimir's energy on 
non-commutative two sphere has been computed. This case differs from the 
others because of the finite number of field modes caused by the 
noncommutativity. So, the Casimir's energy (\ref{ncen}) is finite without 
any renormalization and depends on the radius of the sphere and the 
noncommutative parameter $\ga$. One could expect that an interesting 
thing is to investigate what happens as $\ga$ tends zero. It is very often 
(see for example \cite{ncqm}) 
mentioned that one is able to get the standard 
(commutative) results from the noncommutative ones in the limit $\ga=0$. 
We see that it is not the case when interested in Casimir effect because 
if we do the limit $\ga\to 0$ in (\ref{ncen}) we will not obtain the result 
(\ref{sphres}). We would get only the infinite expression which could be 
understood as the starting point for the renormalization procedure. This 
difference between the features of the Casimir effects on a commutative 
space and its noncommutative version is known in the case of a 
noncommutative cylinder - the first reference of \cite{speef}. 
In this case the spacetime is a 
cylinder times the real time. 
Noncommutativity appears only on the cylinder (the time is commutative). 
The noncommutativity does not lead to a natural cutoff of the field modes 
because the cylinder is not a compact space. So, the computation of 
Casimir's energy involves a renormalization and regularization. The 
result is essentially different from the commutative one. In the case of 
noncommutative cylinder the Casimir's energy per unit length does not 
depend on the cylinder radius $r$ but in the commutative case this 
quantity depends on $r$ as $r^{-2}$. Such discrepances might be caused by 
the fact that our (or \cite{speef}) field theories are not built on the 
space-times with usual symmetries - the time is always added to the 
noncommutative space as an external commuting parameter which cannot be 
mixed with others coordinates. In this way, the situation 
is rather similar to the nonrelativistic quantum mechanics than to a 
relativistic theory. Maybe, it should be found another way to construct 
a field theory on a noncommutative space as presented above or in 
\cite{speef}.

\end{document}